\documentclass[prl,twocolumn,showpacs,preprintnumbers,amsmath,amssymb]{revtex4}

\usepackage[english]{babel}
\usepackage{graphicx}
\usepackage{dcolumn}
\usepackage{bm}
\usepackage{SIunits}

\newcommand{\un}[1]{~\mathrm{#1}}

\begin{document}


\title{Experimental realization of highly-efficient broadband coupling of single quantum dots to a photonic crystal
waveguide }

\author{T.~Lund-Hansen$^1$}
\email{tolh@fotonik.dtu.dk}
\author{S.~Stobbe$^1$}%
\author{B.~Julsgaard$^1$}
\altaffiliation[Now at ]{Physics Department, University of Aarhus,
Denmark.}
\author{H.~Thyrrestrup$^1$}

\author{T.~S\"unner$^2$}
\author{M.~Kamp$^2$}
\author{A.~Forchel$^2$}

\author{P.~Lodahl$^1$}
\email{pelo@fotonik.dtu.dk}
\homepage{http://www.fotonik.dtu.dk/quantumphotonics}

\affiliation{
$^1$DTU Fotonik, Department of Photonics Engineering, Technical University of Denmark, \O rsteds Plads 343, DK-2800 Kgs. Lyngby, Denmark\\
$^2$Technische Physik, Universit\"at W\"{u}rzburg, Am Hubland, D-97074 W\"{u}rzburg, Germany
}%
\date{\today}
\begin{abstract}
We present time-resolved spontaneous emission measurements of
single quantum dots embedded in photonic crystal waveguides.
Quantum dots that couple to a photonic crystal waveguide are
found to decay up to $27$ times faster than uncoupled quantum dots. From these measurements $\beta$-factors of up to 0.89 are
derived, and an unprecedented large bandwidth of 20 nm is
demonstrated. This shows the promising potential of photonic crystal waveguides for efficient single-photon sources. The scaled
frequency range over which the enhancement is observed is in excellent
agreement with recent theoretical proposals taking into account that the light-matter
coupling is strongly enhanced due to the significant slow-down of
light in the photonic crystal waveguides.
\end{abstract}

\pacs{78.47.-p, 42.50.Ct, 78.67.Hc}

\maketitle

The ability to control light-matter dynamics using photonic
crystals has been demonstrated experimentally within the last few
years \cite{Kress2005, Englund2005, Hennessy2007, Noda2007,
Lodahl2004}. A particularly attractive application of photonic
crystals is on-chip single-photon sources. A highly efficient
single-photon source is the key component required in many quantum
communication protocols \cite{Lounis2005} and will allow
implementing linear optics quantum computing \cite{Knill2001}.
Single photons are harvested when a quantum dot (QD) is coupled
efficiently to an enhanced optical mode. One very successful
approach has been to couple single QDs to a photonic
crystal nanocavity \cite{Kress2005, Englund2005, Hennessy2007}. In
this process, photons are emitted with large probability to the
\emph{localized mode} of the cavity at a strongly enhanced rate.
One drawback of this approach, however, is that the photons
subsequently must be coupled out of the cavity, which will reduce
the overall efficiency of the device significantly. Furthermore,
nanocavity single-photon sources only operate within a narrow
bandwidth determined by the high Q factor of the cavity. Very
recently it was proposed that photonic crystal waveguides (PCWs)
provide a way of overcoming these limitations \cite{Hughes2004,Rao2007a,
Rao2007, Lecamp2007}, which was inspired by early work of Kleppner
on metallic waveguides \cite{Kleppner1981}. Here we present
the experimental verification that single QDs can be coupled
efficiently to the mode of a PCW.

PCWs offer the possibility to tailor the
dispersion of light by proper design of the structure. In this way
impressive light slow-down factors of 300 have been experimentally
demonstrated \cite{Vlasov2005}. The efficient slow-down of the
PCW mode implies that the light-matter coupling strength
will be largely enhanced. This enhanced coupling will allow the
efficient channeling of single photons from a QD into the
PCW mode. In this case,
the photons are transferred directly to the \emph{propagating
mode} of the PCW, which is fundamentally different from the
cavity case, and implies that the overall efficiency of the source
is potentially very high. Furthermore, the enhancement in a
PCW is not limited to a narrow spectral bandwidth as in a cavity, and precise control over the QD position is not required. Consequently demands for spatial and spectral tuning of the emitter are less stringent for PCWs than for photonic crystal nanocavities \cite{Hennessy2007}.

\begin{figure}[t]
\includegraphics[width=\linewidth]{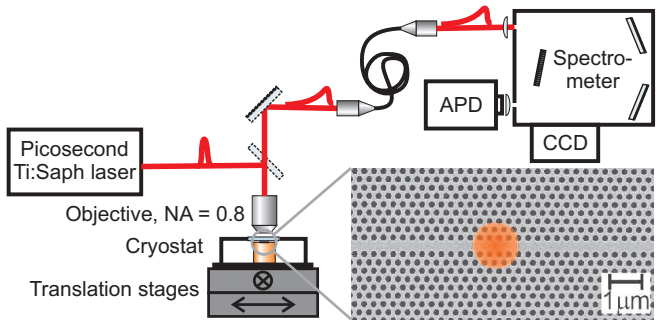}
\caption{\label{fig:setup} (Color online) Sketch of the experimental
setup, which is described in detail in the text. The inset to the
bottom right is a scanning electron micrograph of a fabricated
PCW. The red sketched region illustrates
the size of the area from which spontaneous emission is collected.}
\end{figure}

In this Letter, we present time-resolved spontaneous emission
measurements on single QDs positioned in PCWs. The PCWs are formed by leaving out a single row
of holes in the triangular lattice of the photonic crystal, see
inset of Fig.~\ref{fig:setup}.  The photonic crystals are
fabricated using electron beam lithography followed by dry and wet
etching. In this way $150 \un{nm}$ thick GaAs membranes are
obtained containing a single layer of self-assembled InAs QDs at the center. The density of QDs is $\sim 250
\un{\micro m}^{-2}$ with a ground state emission wavelength
centered at $960 \un{nm}$ and inhomogeneously broadened with a
width of $50 \un{nm}$. Two PCW samples have been fabricated,
one with lattice parameter $a=248 \pm 2 \un{nm}$ and radius
$r=(0.292\pm 0.006)a$ and one with $a=256 \pm 2 \un{nm}$ and
$r=(0.286\pm 0.006)a$. The samples are $17 \un{\micro m}$ wide and $100 \un{\micro m}$ long such that finite size effects can be neg\-lected. 

The experimental setup is shown in Fig.~\ref{fig:setup}. The
QDs are optically excited with a pulsed Ti:Sapphire laser
operating at $800\un{nm}$ with a pulse length of $\sim 2 \un{ps}$
and a repetition rate of $75 \un{MHz}$. Excitation is done through
a high numerical aperture microscope objective ($NA = 0.8$) in
order to limit the contribution from QDs outside the
PCW. The fraction of the spontaneously emitted photons that
couple to radiation modes are collected through the same
microscope objective and subsequently focussed to a single mode
fiber that acts as a confocal pinhole for spatial selection. The
spatial resolution is found to be $1.4 \un{\micro m}$ (cf.
diameter of collection area indicated in Fig.~\ref{fig:setup}) by
determining the distance over which the sample can be moved
relative to the collection optics before the emission from a
single QD is halved. The spatially filtered spontaneous
emission is coupled to a spectrometer equipped with a CCD for
recording spectra (spectral resolution $0.15\un{nm}$) or an
avalanche photo diode (APD) for time-correlated single-photon
spectroscopy measurements (temporal resolution $280 \un{ps}$). The
sample is placed in a helium flow cryostat operating at $10
\un{K}$. The cryostat is mounted on high-precision motorized
translation stages, which enables positioning of the sample with a
precision of $100 \un{nm}.$

\begin{figure}[t]
\includegraphics[width=\linewidth]{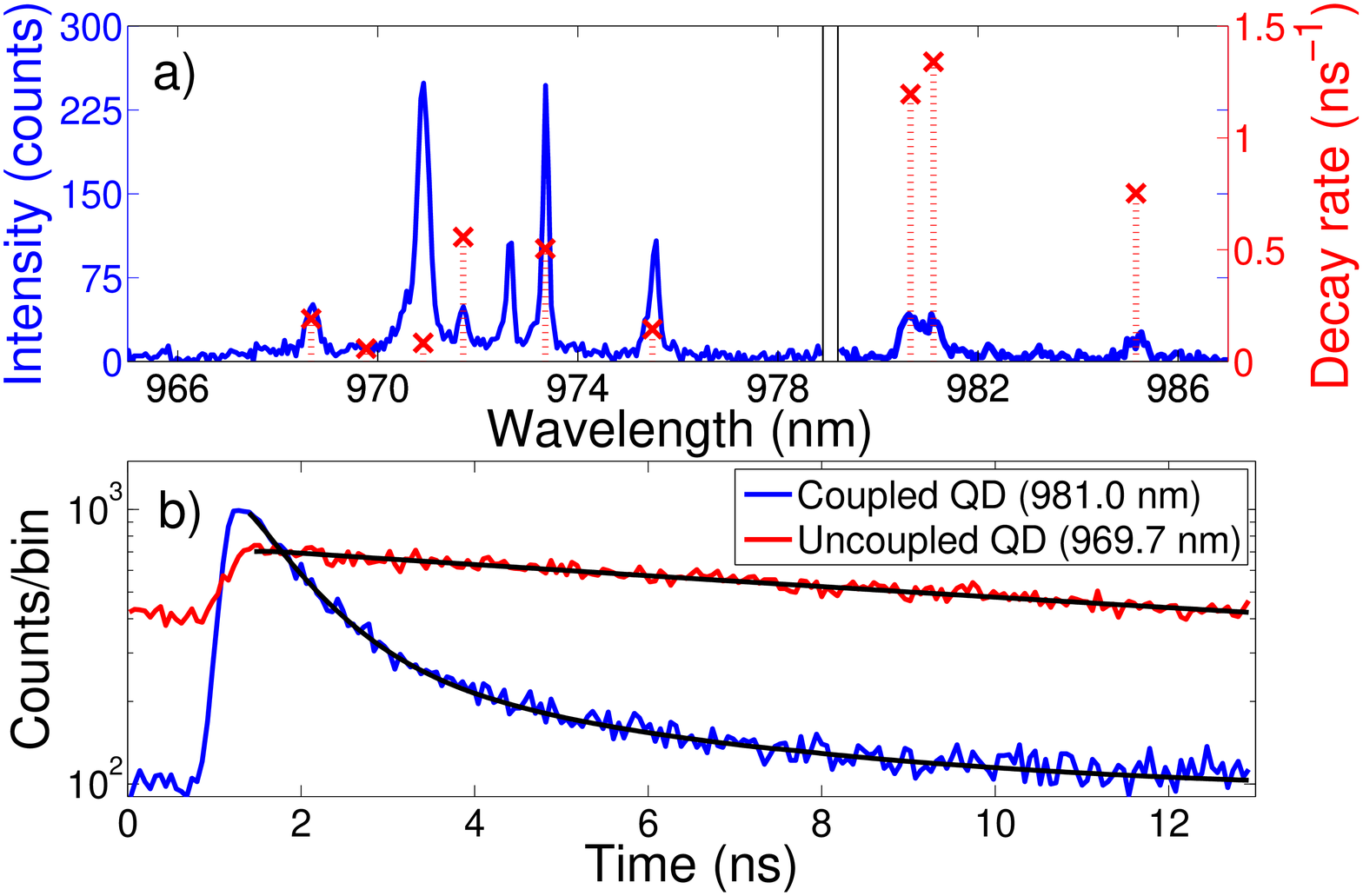}
\caption{\label{fig:spectrum_decay} (Color online)
(a) Spontaneous emission spectra (solid blue curve) displaying single QD
emission lines. The spectra were recorded at two different
positions on a PCW with $a = 256 \un{nm}$.  The excitation density was $\sim 3 \un{W/cm^2}.$
The crosses indicate the measured decay rates (right axis) of the
different QD emission lines. (b) Decay curves from two single QD lines. The blue decay curve is
measured at $981.0 \un{nm}$ where the QD is coupled to the
PCW. The red decay curve is measured at
$969.7 \un{nm}$ and is an example of an uncoupled QD. The
black lines are the fitted single (red curve) or double (blue
curve) exponential decay models.}
\end{figure}

Examples of spontaneous emission spectra recorded at two different
positions on a PCW ($a = 256 \un{nm}$) are
displayed in Fig. \ref{fig:spectrum_decay}. We observe discrete
single QD emission lines. The measurements are carried
out in the low excitation regime below the saturation level of
single exciton lines, where predominantly spontaneous emission
from the QD ground state is observed. The varying heights
of the emission peaks reflect that the QD emission is
redistributed depending on its position in the photonic crystal
membrane \cite{Kaniber2008}. In general we observe low emission
from QDs that couple to the PCW.
Clear spectral signatures of QD emission coupled to a
PCW performed in a transmission geometry
are reported in \cite{Viasnoff-Schwoob2005a}. Here we employ
time-resolved spontaneous emission measurements as a way to
directly determine the coupling rate of photons from the QD to the PCW.

Time-resolved spontaneous emission has been recorded on single
QD lines at a large number of different emission
wavelengths. Two examples of decay curves that reveal very
different QD dynamics are displayed in
Fig.~\ref{fig:spectrum_decay} (b). One of the decay curves
displays a very slow single exponential decay with a rate of $0.05
\un{ns^{-1}}$. This decay curve corresponds to a QD that
is not coupled to the PCW, which can be due to spatial
mismatch relative to the PCW or that the QD dipole
moment is oriented along the PCW axis \cite{Lecamp2007}. The
decay rate of the uncoupled QD is inhibited by a factor
$\sim 20$ compared to a QD in a homogeneous medium, which
is an effect of the 2D photonic bandgap of the photonic crystal
membrane \cite{Kaniber2008,Koenderink2006}. The decay from the other QD in Fig.~\ref{fig:spectrum_decay} (b) is much faster due to coupling to the PCW mode. The fast
decay curve is well described by a double exponential model, where
the fast component is the rate due to coupling to the PCW,
while the slow component contains contributions from dark exciton
recombination in the QD \cite{Smith2005, Johansen2008}.
Note that a weak slow component also will contribute to the decay
curve of the uncoupled QD, however in this case it cannot
be distinguished from the inhibited decay, and a
single-exponential model is sufficient. For the fast decay curve
displayed in Fig.~\ref{fig:spectrum_decay} (b) we derived a fast
decay rate of the coupled QD of $1.34 \un{ns^{-1}}$.
Consequently the coupled QD decays a factor of $27$ times
faster than the uncoupled QD, which demonstrates that
photons can be channelled very efficiently into the PCW, in agreement with recent theoretical proposals \cite{Rao2007a, Rao2007, Lecamp2007}.

\begin{figure}[t]
\includegraphics[width=\linewidth]{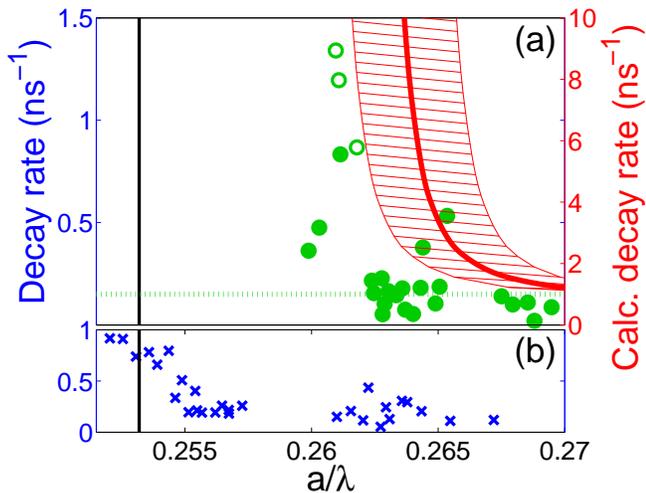}
\caption{\label{fig:a256decay_vs_al}(Color
online) (a) Measured QD decay rates on the PCW sample with $a = 256 \un{nm}$ as a function of
scaled frequency. The excitation power for these measurements was
$3 \un{W/cm^2}$. The data points marked by open (filled) circles
are modeled using a double (single) exponential decay model. The
red line (right axis) displays the decay rates calculated from numerical
simulations with uncertainties in $a$ of $\pm 2 \un{nm}$ given by
the hatched area. The black line at $a/\lambda = 0.253$ is the
calculated band edge of the 2D photonic bandgap. The dotted green
line is the mean decay rate of uncoupled QDs used for the
estimation of the $\beta$-factor. (b) Measured decay rates on a photonic
crystal sample with no PCW ($a = 256 \un{nm}$).}
\end{figure}

The complete set of in total $26$ measured decay rates on the
sample with $a = 256 \un{nm}$ is shown in Fig.~\ref{fig:a256decay_vs_al} (a). Here the measured decay rates are
plotted as a function of scaled frequency $a/\lambda$, where
$\lambda$ is the emission wavelength. In modeling the data, in
most cases single-exponential decay curves suffices, while only
for the fastest decay curves a bi-exponential model was needed.
The single-exponential model was abandoned when the properly
normalized sum of the residuals characterizing the fit, i.e.~the
reduced $\chi^2$, was above $1.3$. We observe a range of decay
rates since differently positioned and oriented QDs
couple differently to the PCW, as discussed
above. Fast decay rates are only observed for a limited range of
$a/\lambda$, which is in very good agreement with expectations
from theory. Hence the most efficient coupling to the PCW occurs when the QD can couple to a
slowly propagating mode, i.e.~when the emission wavelength is
tuned to the edge of the PCW dispersion relation.

To compare the frequency dependence of the measured decay rates
with theory we have calculated the decay rate of a QD
positioned in the center of the PCW using
the theory of \cite{Rao2007a}:
\begin{eqnarray}
\Gamma=\Gamma_0 \frac{3 \pi c^3 a}{V_{\mathrm{eff}}\omega^2\varepsilon^{3/2}v_g(\omega)},
\end{eqnarray}
where $\omega$ is the frequency of the emitter, $c$ is the speed
of light in vacuum, and $\varepsilon$ is the electric
permittivity. The group velocity ($v_g$) and effective mode volume
($V_{\mathrm{eff}}$) have been extracted from a band structure
calculation using the MPB software package \cite{Johnson2001:mpb}.
$\Gamma_0$ is the decay rate in a homogeneous material, which is
$1.1 \un{ns^{-1}}$ for the QDs in this experiment. In
Fig.~\ref{fig:a256decay_vs_al} (a) we have plotted the resulting
calculated decay rates (red line) assuming no adjustable
parameters. A discrepancy of only 1~\% is found between the
frequency of enhanced decay rates in the measurement and the point
of divergence of the calculated decay rate. The main uncertainty
in the calculated decay rate originating from the uncertainty in
$a$ is illustrated in Fig.~\ref{fig:a256decay_vs_al} (a) by the
hatched area. Anticipating the additional uncertainty in $r$ and
$\epsilon$, and the numerical uncertainty of the calculation, we
conclude that there is an excellent match regarding the range of
scaled frequencies where enhanced rates are observed. This clearly
proves that the enhancement is due to coupling to the PCW. 
%
Our measured decay rates are found to be approximately 8 times smaller than the calculated decay rates. A lower decay rate is expected as the theory assumes a dipole emitter positioned and oriented optimally with respect to the
PCW and does not take into account imperfections giving rise to scattering losses. These are known to limit the achievable group velocity
slow-down factor thereby removing the divergence of the decay rate \cite{Hughes2005}. Note that scattering loss is mainly a limitation for devices relying on long propagation distances, while a single-photon source can be made very compact, thus making the use of slow-light in a photonic crystal waveguide a viable approach.

Shown in Fig.~\ref{fig:a256decay_vs_al} (b) are the measured
QD decay rates in a photonic crystal membrane $(a = 256
\un{nm})$ without a PCW. This is done in order to locate the
edge of the 2D photonic bandgap of the photonic crystal in order
to ensure that the enhancement discussed above is not an effect of
the band edge. We observe an increase in the decay rate due to the
band edge at $a/\lambda=0.254$, which distinctly differs from the
scaled frequency where the PCW coupling is
observed. Furthermore, the position of the band edge matches the
value found from the band structure calculation of 0.253 very well
(marked by the black line in Fig.~\ref{fig:a256decay_vs_al}).
Across the band edge the QD decay rates are observed to
increase to around $1 \un{ns^{-1}}$ while inside the 2D band gap
decay rates between $0.05 \un{ns^{-1}}$ and $0.43\un{ns^{-1}}$ are
observed. The observed fluctuations reflect the dependence of the
projected local density of optical states on the QD
orientation and position. The measurements can be compared to the
calculations of Koenderink \emph{et al.} \cite{Koenderink2006},
where inhibition factors between 0.03 and 0.39 are predicted in
the respective energy range of the 2D band gap.

\begin{figure}[t]
\includegraphics[width=\linewidth]{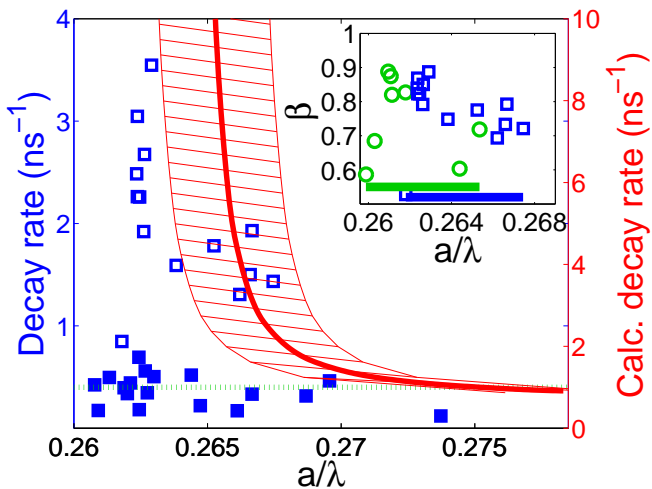}
\caption{\label{fig:a248decay_vs_al}(Color
online) Measured decay rates for QDs in a PCW sample with $a = 248 \un{nm}.$ The pump
excitation power was $1.5 \un{kW/cm^2}$. The data points marked by
open (filled) squares are modelled using a double (single)
exponential decay model. The red line (right axis) shows the decay rates
calculated from numerical simulations with uncertainties in $a$ of
$\pm 2 \un{nm}$ given by the hatched area. The dotted green line
is the mean decay rate of uncoupled QDs used for the
estimation of $\beta$. The inset displays the $\beta$-factors
above 0.5 calculated from the data in
Fig.~\ref{fig:a256decay_vs_al} (green circles) and
Fig.~\ref{fig:a248decay_vs_al} (blue squares) as described in the
text. The green (blue) bar shows the coupling bandwidth of
the PCW sample with $a=256\un{nm}$ ($a=248\un{nm}$).}
\end{figure}

We have collected further experimental evidence for our
conclusions by investigating another PCW on
a sample with lattice parameter $a=248\un{nm}$.  In this case the
slow light regime, where efficient coupling to the PCW
occurs, matches the excited state of the QDs.
Consequently these measurements were performed in the highly
saturated regime where spontaneous emission from the QD
excited states is observed. The data are presented in
Fig.~\ref{fig:a248decay_vs_al}. Once again the data fall into two
groups, in this case below and above $\sim 0.5 \un{ns}^{-1}$, of
slow and fast rates corresponding to uncoupled and coupled QDs, respectively. The rates of the uncoupled QDs are
relatively fast, since the excited states are known to have
increased non-radiative decay compared to the ground state
excitons. Strong enhancement is observed in this case for
$a/\lambda=0.263$, which matches theory very well. We observe
enhanced rates of up to $3.5 \un{ns}^{-1}$ clearly demonstrating
the very pronounced effect of the PCW.

The figure of merit determining the coupling efficiency into the
PCW is the $\beta$-factor. It is defined as \cite{Rao2007a}
\begin{eqnarray}
\beta= \frac{\Gamma_{\mathrm{wg}}}{\Gamma_\mathrm{{wg}}+\Gamma_{\mathrm{rad}}+\Gamma_{\mathrm{nr}}},
\end{eqnarray}
where $\Gamma_{\mathrm{wg}}$ is the decay rate of the QD
to the PCW, $\Gamma_{\mathrm{rad}}$ is the radiative decay
rate to non-guided modes, and $\Gamma_{\mathrm{nr}}$ is the
intrinsic QD non-radiative decay rate
\cite{Johansen2008}. $\Gamma_{\mathrm{tot}} \equiv
\Gamma_{\mathrm{rad}}+\Gamma_{\mathrm{nr}}$ can be extracted from
the measurements on QDs that do not couple to the
PCW. The decay rate to non-guided modes will depend on
position and orientation of the individual QD
\cite{Lecamp2007,Koenderink2006}, which is reflected in the
variations in the decay rates of the uncoupled QDs in
Fig.~\ref{fig:a256decay_vs_al} and \ref{fig:a248decay_vs_al}. To
accommodate this we extract the average total decay rate of the
uncoupled QDs, which is $\Gamma_{\mathrm{tot}}(a=248
\un{nm}) = 0.4 \un{ns^{-1}}$ and $\Gamma_{\mathrm{tot}}(a=256
\un{nm}) = 0.15\un{ns^{-1}}$ for the two data set, respectively
(marked by the green dotted line in Fig.~\ref{fig:a256decay_vs_al}
and Fig.~\ref{fig:a248decay_vs_al}). In the inset of
Fig.~\ref{fig:a248decay_vs_al} the $\beta$-factor is plotted
versus scaled frequency. We observe $\beta$-factors of up to 0.89,
demonstrating the excellent photon collection efficiency of
PCWs. Even more spectacularly, a
$\beta$-factor above 0.5 is observed in a relative bandwidth as
large as 2~\% (corresponding to $20 \un{nm}$) for both PCW
samples. This superior bandwidth is unique for a PCW. For comparison a $\beta$-factor of 0.92 limited to a
relative bandwidth of 0.3~\% has been demonstrated in photonic
crystal cavities \cite{Chang2006}. This demonstrates the important
advantage of PCWs for high-efficiency large
bandwidth single-photon sources.

We have experimentally demonstrated that spontaneous emission from
single QDs can be coupled very efficiently to a PCW. The light-matter coupling is enhanced by the
light slow-down mediated by the dispersion control provided by the
PCW. A $\beta$-factor as high as 0.89 and an unprecedented large bandwidth of $20\un{nm}$ has
been obtained,
in this respect outperforming the traditional QD
single-photon source approach based on narrow bandwidth cavities.

We thank J\o rn M. Hvam for fruitful discussion. We gratefully
acknowledge the Danish Research Agency for financial support
(projects FNU 272-05-0083, 272-06-0138 and FTP 274-07-0459). BJ is
supported by The Carlsberg Foundation.


\end{document}